\documentclass[useAMS,usenatbib]{mn2e}

\usepackage{graphicx}

\newcommand{\ion}[2]{#1~{\sc #2}}

\title[New CP stars with resolved magnetically split lines]{
Two new chemically peculiar stars with resolved magnetically split lines\thanks{Based on observations 
obtained at the European Southern
Observatory, La Silla, Chile (ESO programme 077.D-0477(A)).}}
\author[S. Hubrig \& N. Nesvacil]{S. Hubrig$^{1}$\thanks{E-mail: shubrig@eso.org,},
N. Nesvacil.$^{2}$ \\
$^{1}$European Southern Observatory, Casilla 19001, Santiago, Chile\\
$^{2}$Department of Astronomy, University of Vienna, T\"urkenschanzstrasse 17, 1180 Vienna, 
Austria
\\
}

\begin{document}

\date{Accepted 2007 Enero 99. Received 2007 Enero 98}

\pagerange{\pageref{firstpage}--\pageref{lastpage}} \pubyear{2007}

\maketitle

\label{firstpage}

\begin{abstract}
We report the discovery of resolved magnetically split lines in two chemically peculiar stars,
the SrEuCr star HD\,92499 and the Bp SiCr star HD\,157751.
From FEROS spectra, we have measured a mean magnetic field modulus of 8.5\,kG for HD\,92499 and 
a  mean magnetic field modulus of 6.6\,kG for HD\,157751.
Both stars have small projected rotational velocities: $v \sin i$ = 3.0~km$\,$s$^{-1}$ and 
8.5~km$\,$s$^{-1}$, respectively. Our preliminary abundance analysis reveals ionisation imbalance of 
rare earths in HD\,92499 indicating the abundance pattern typical of rapidly oscillating Ap stars.
Cr and Fe are found strongly overabundant in HD\,157751.

\end{abstract}


\section{Introduction}
\label{sect:intro}
Between 2003 and 2006 we have been carrying out 
a systematic search for longitudinal magnetic fields in
chemically peculiar stars whose magnetic field has never been studied before
\citep[e.g.,][]{2006AN....327..289H,2005ASPC..343..374H}.
The goal of this study was to statistically enlarge our data sample
by including all southern stars for which the position in the
H-R diagram is known from accurate Hipparcos \citep{ESA97} parallaxes 
($\sigma(\pi)/\pi<0.2$) and
from photometric data in the Geneva and Str\"omgren systems, used 
to determine their effective temperatures.
The longitudinal magnetic field determinations have been obtained with FORS\,1 at the VLT, which is a multi-mode 
instrument equipped with polarisation analyzing optics comprising super-achromatic half-wave and 
quarter-wave phase retarder plates, and a Wollaston prism with a beam divergence of 22$\arcsec$ 
in standard resolution mode. We used the GRISM\,600B and the GRISM\,1200g to cover several 
hydrogen Balmer lines at a spectral resolution of R$\sim$2000 for the GRISM\,600B and R$\sim$4000 
for the GRISM\,1200g. A few stars with detected strong longitudinal magnetic fields 
have been successively scheduled for the observations with high-resolution spectrographs.
In the course of such a study we had discovered a very slowly rotating extreme magnetic Ap star,
HD\,154708 (=\,CD\,$-$57\degr6753), which has the second highest
value of a field modulus ever measured in Ap and Bp stars \citep{2005A&A...440L..37H}.
Up to now, the number of magnetic stars with a measured
mean magnetic field modulus using magnetically resolved lines was 49.
In this paper we present our most recent discovery of two additional strongly magnetic slowly rotating 
chemically peculiar 
stars, the  SrEuCr star HD\,92499 (=\,CD\,$-$42\degr6407) and the Bp SiCr star HD\,157751 (=\,CD\,$-$33\degr12069), for which 
the Zeeman pattern is resolved for several spectral lines. 

\section{Observations and magnetic field measurements.}
\label{sect:obs}

\begin{figure}
\centering
\includegraphics[width=0.45\textwidth]{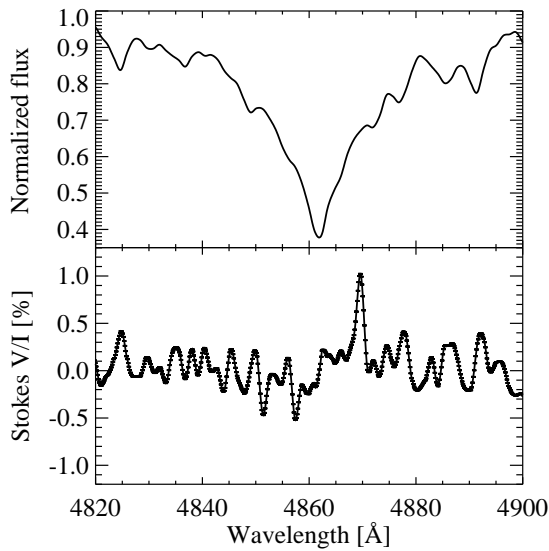}
\caption{
Stokes I and V/I spectra of the Ap star HD\,92499 in the vicinity of the H$\beta$ line.
}
\label{fig:IV92499}
\end{figure}

\begin{figure}
\centering
\includegraphics[width=0.45\textwidth]{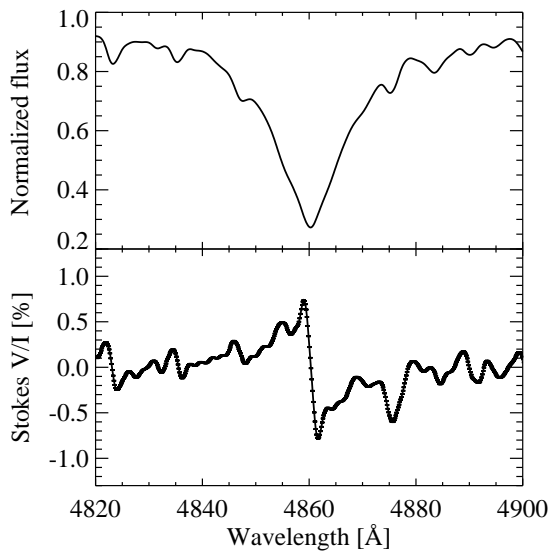}
\caption{
Stokes I and V/I spectra of the Ap star HD\,157751 in the vicinity of the H$\beta$ line.
}
\label{fig:IV157751}
\end{figure}

The observations of HD\,92499 and HD\,157751
have been carried out with the echelle spectrograph FEROS at the 2.2\,m  telescope at La Silla.
The spectrum of HD\,92499 has been obtained on May 13, 2006, and the spectrum of 
HD\,157751 on June 14, 2006. The S/N ratio of the spectra is about 
200--300 per pixel in the one-dimensional spectrum around 
6150\,\AA{}.
The wavelength coverage is 
$3530-9220$\,\AA{}, and the nominal resolving power is 48\,000.
To our knowledge, these observations are the first 
carried out for both stars at high spectral resolution. 
Similar to the recently discovered cool
low mass Ap star HD\,154708, HD\,92499 and HD\,157751 have never been studied 
in detail and the SIMBAD database offers merely a few references.
As mentioned above, the only studies of 
both stars have been 
conducted with the purpose of longitudinal magnetic field determination with FORS\,1 at the VLT.
The longitudinal magnetic fields have been determined from the measurements of the circular 
polarisation of opposite sign induced in the wings of 
each Balmer line by the Zeeman effect
\citep[see][for more details]{Hu04a,Hu04b}.
Our measurements using all hydrogen Balmer lines from 
H$\beta$ to the Balmer jump at a spectral resolution of R$\sim$2000 revealed  for both stars 
kilogauss longitudinal magnetic
fields $\left<B_z\right>$ of the order of $-$1.2\,kG for HD\,92499 and 4.0\,kG for 
HD\,157751 \citep{2006AN....327..289H}. 
In Figs.~\ref{fig:IV92499} and \ref{fig:IV157751} we present both, the Stokes I and V/I 
spectra of these stars in the vicinity of the H$\beta$ line. 

The visual inspection of high resolution FEROS spectra disclosed that in both stars
spectral lines are strongly affected by the magnetic field. Several spectral line 
profiles observed in HD\,92499 are fully resolved whereas for HD\,157751 the magnetic splitting 
is less noticeable due to non-negligible rotational Doppler effect. From the observed 
magnetically split lines we determined the mean magnetic field modulus (that is the average 
over the visible stellar hemisphere of the modulus of the magnetic vector, weighted by 
the local line intensity) which is related to the wavelength separation of the Zeeman components 
through the relation 
\begin{equation}
\centering
\left<B\right>=\Delta\lambda/(9.34\cdot10^{-13}\;\lambda_{c}^{2}\;g_{\rm eff})
\end{equation}
where $\lambda_{\rm c}$ is the central wavelength of the line 
(in a Zeeman triplet this corresponds to the position of the $\pi$-component), ${\rm \Delta}\lambda$ is the wavelength separation between the 
centroids of the $\sigma$-components and 
$g_{\rm eff}$ is the effective  Land\'e  factor. 
To calculate ${\rm \Delta}\lambda$, the wavelengths of the centres of gravity of the
split doublet and triplet components have been determined by fitting a gaussian 
simultaneously
to each of them (if the lines are not fully split) or by direct integration of
the whole component profile (if the splitting is large). When the lines were blended,
a multiple fit of three or four gaussians has been performed.
This procedure is well-established over the years and has been used in previous studies 
of magnetic stars with resolved magnetically split lines \citep[e.g.,][]{mat97}. 
In this manner we obtained the mean magnetic field modulus 
$\left<B\right>$ = $8.5\pm0.2$\,kG for HD\,92499  and $\left<B\right>$ = $6.6\pm0.5$\,kG
for HD\,157751.
The accuracy of our measurements essentially depends on the magnitude of the Zeeman 
splitting, presence of blends and the complexity of the profiles. 

Strongly magnetic stars with measured mean magnetic fields from magnetically resolved 
lines are of special
interest because they provide the best opportunity to study the effect of the magnetic
field on the stellar atmospheres.
Very frequently, abundances of many elements are non-uniformly distributed
over the stellar surface and they are also
vertically stratified.
To get more insight in the peculiar nature of HD\,92499  and HD\,157751, we  
tried to determine abundances of a few elements

To calculate the effective temperature  $T_{\rm eff}$ and $\log g$ we used  $uvby\beta$ and
Geneva photometry listed in the General Catalogue of Photometric Data \citep{mer97}.
Further we made use of the software package TempLogG \citep{ro95,st02} 
which offers the possibility to automatically correct for the reddening, 
to obtain $T_{\rm eff}=7000$\,K, $\log g=4.15$ 
for HD\,92499, and $T_{\rm eff}$= 11380\,K, $\log g$=4.44 for HD\,157751.
The calculations of abundances for a few lines - mainly for Fe and Cr lines - were 
made with an evolved version of the WIDTH9 code \citep{kur05}.
This code takes into account 
the extra broadening induced by the magnetic field 
and gives a good first estimate which can later be refined by more detailed analyses using 
spectral synthesis.
Iteratively, the dependence of the average Fe abundance on excitation potentials of the 
measured lines was minimised to derive a more accurate value of $T_{\rm eff}$.

The best fit to the observed spectrum of HD\,92499 has been achieved for 
$T_{\rm eff}$ = $7200\pm200$\,K, and  $\log g$ = $4.15\pm0.15$.
For HD\,157751, we obtained 
$T_{\rm eff}$ = $11300\pm300$\,K, and  $\log g$ = $4.4\pm0.2$.
Our estimates of the projected velocity  $v$\,$\sin i$ using several Fe and Cr lines with small 
Land\'e factors resulted in 
$v$\,$\sin i$ = $3.0\pm0.5$\,km\,s$^{-1}$
for HD\,92499 and $v$\,$\sin i$ = $8.5\pm1.0$\,km\,s$^{-1}$
for HD\,157751.
Atomic line data were taken from VALD \citep{ku99}. Iteratively, the derived Fe and Cr abundances were then used to calculate a model with individual abundances using LLmodels code \citep{sh04}.
Atmospheric parameters and abundances of \ion{Fe}{ii} and \ion{Cr}{ii} model calculations are presented in Table~\ref{tab:results}.

Employing the calculated model for HD\,92499, we have determined abundances for a few rare earth elements 
using strong unblended lines. Europium, praseodymium and neodymium 
have been found strongly overabundant, as typical for most magnetic chemically peculiar stars.
The computed respective relative abundances are: 
$\log(N_{\rm EuII}/N_{\rm tot})=-9.41$, 
$\log(N_{\rm NdII}/N_{\rm tot})=-8.15$, 
$\log(N_{\rm NdIII}/N_{\rm tot})=-7.29$,
$\log(N_{\rm PrII}/N_{\rm tot})=-9.31$,
$\log(N_{\rm PrIII}/N_{\rm tot})=-8.13$ 
with estimated uncertainties of 0.3 dex. Interestingly, we find the obvious imbalance of Nd and Pr
abundances from different ions, also called as "rare-earth-anomaly". The abundances derived from the first 
and second ions differ by $\sim$0.86~dex for Nd and by $\sim$1.18~dex for Pr. Such an anomaly is usually  
explained by vertical abundance stratification of the elements in the
stellar atmosphere and is known to be a typical feature of the 
abundance pattern of the rapidly oscillating Ap (roAp) stars \citep{Ry04}. It is generally accepted that 
the vertical abundance stratification is a consequence of atomic diffusion, i.e. radiative 
levitation and gravitational settling. 

\begin{figure}
\centering
\includegraphics[height=0.45\textwidth,angle=270]{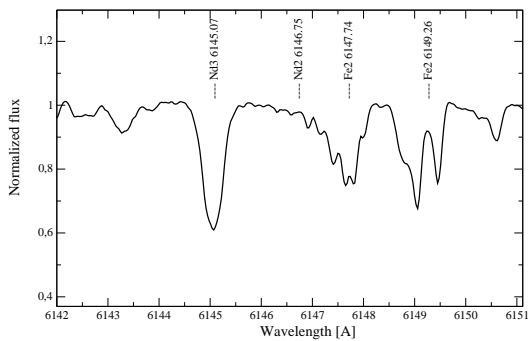}
\caption{
The spectrum of HD\,92499 in the spectral region
around the \ion{Fe}{ii} lines $\lambda$\,6147.7 and $\lambda$\,6149.2.
}
\label{fig:hd92499_6149_onlyobs}
\end{figure}

\begin{figure}
\centering
\includegraphics[height=0.45\textwidth,angle=270]{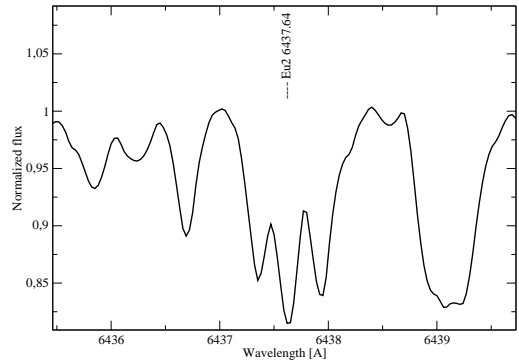}
\caption{
The spectrum of HD\,92499 in the spectral region
around the \ion{Eu}{ii} line $\lambda\,6437.6$.
}
\label{fig:HD92499_Eu2_onlyobs}
\end{figure}

\begin{figure}
\centering
\includegraphics[height=0.45\textwidth,angle=270]{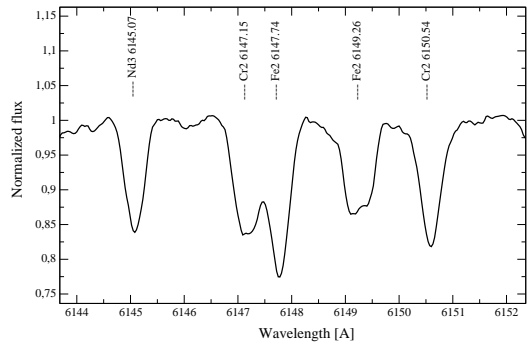}
\caption{
The spectrum of HD\,157751 in the spectral region
around the \ion{Fe}{ii} lines $\lambda\,6147.7$ and $\lambda\,6149.2$.
}
\label{fig:HD157751_Fe6149_onlyobs}
\end{figure}

\begin{figure}
\centering
\includegraphics[height=0.45\textwidth,angle=270]{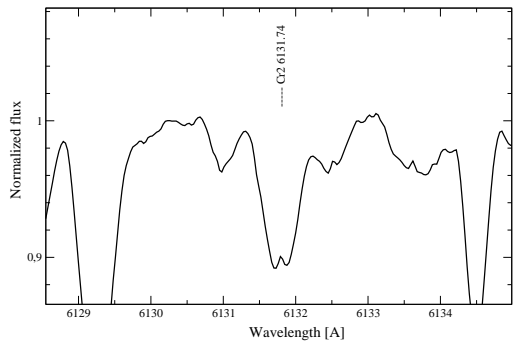}
\caption{
The FEROS spectrum of HD\,157751 in the spectral region
around the \ion{Cr}{ii} line $\lambda\,6131.7$.
}
\label{fig:hd157751_6130-6133_onlyobs}
\end{figure}

The observed FEROS spectra and line identification according to VALD in 
various spectral regions containing magnetically split lines are presented 
for both stars in Figs.~\ref{fig:hd92499_6149_onlyobs}--\ref{fig:hd157751_6130-6133_onlyobs}.

\begin{table}
\caption{Atmospheric parameters and abundances of \ion{Fe}{ii} and \ion{Cr}{ii} used for 
spectrum synthesis.}
\label{tab:results} 
\begin{center}
\begin{tabular}{l | c c c}
\hline
                                   &HD~92499    &HD~157751\\
\hline
$T_{\rm{eff}}$ [K]                 &$7200\pm200$   & $11300\pm300$ \\
${\rm{log}}~g$                     &$4.15\pm0.15$  & $4.4\pm0.2$ \\
$v~{\rm{sin}}~i$ [$\rm{km~s^{-1}}$]&$3\pm0.5$      & $8,5\pm1$    \\
$\left<B\right>$ [kG]              &$8.5\pm0.2$    &$6.6\pm0.5$    \\
${\rm log}(N_{\rm FeII}/N_{\rm tot})$ &$-$4.28 &   $-$3.25\\
${\rm log}(N_{\rm CrII}/N_{\rm tot})$ &$-$5.40 &   $-$3.31\\
\hline
\end{tabular}
\end{center}
\end{table}

\section{Discussion.}
\label{sect:disc}

Strong magnetic stars with low $v\,\sin i$ values, for which
the magnetic splitting of the spectral lines is evident in high resolution spectra
in unpolarised light, present an excellent opportunity to derive stronger 
constraints on the geometry of their magnetic fields from the observations of 
the variability of the magnetic field modulus and the longitudinal 
magnetic field over the rotation cycle. Further information about the magnetic field geometry 
can be obtained from the study of the field moments measured in the circular polarisation 
profiles and in line profiles observed in unpolarised light as it has been done 
by \citet{la00}.
Additional observations of these two stars are needed to allow to determine 
their rotational periods, the magnetic field geometry and to carry out 
a multi-element abundance analysis.

Our preliminary abundance analysis of the FEROS spectra of HD\,92499 and HD\,157751
reveals the overabundance of Cr in both stars, overabundance of Fe in HD\,157751 and 
a nearly solar abundance of Fe in HD\,92499.
The discovery of ionisation anomalies of Nd and Pr in HD\,92499 indicates the
abundance pattern typical of roAp stars. 
Because of the location of HD\,92499 in the same region of parameter
space in which rapidly oscillating Ap stars have been detected, this star
was searched photometrically for pulsations by \citet{mar94} with no detected oscillations.
However, we recall the recent paper by \citet{ku06} reporting the discovery of  
the presence of a single frequency of $\nu = 2.088$\,mHz ($P = 8.0$\,min) with very low 
amplitudes in the range $30 - 60$\,m\,s$^{-1}$ from a high-precision radial
velocity study of HD\,154708. No oscillations had previously been 
detected in this star in photometric studies.
We conclude that a time resolved high resolution spectroscopic study of radial velocity 
is called for to confirm or withdraw  the presumption that  HD\,92499 could be a low-amplitude 
roAp star for which photometry would not detect oscillations.

\section*{Acknowledgments}
\label{sect:ackn}
This work was supported by the Austrian Science Fund (FWF-PP17890).


\label{lastpage}

\end{document}